\documentstyle[11pt,epsfig]{article}
\newcommand \be{\begin{equation}}
\newcommand \bea{\begin{eqnarray}}
\newcommand \ee{\end{equation}}
\newcommand \eea{\end{eqnarray}}

\begin{document}

\title{The \$-Game}

\author{J\o rgen Vitting Andersen$^{1,2}$ and Didier Sornette$^{2,3}$\\
$^1$ U. F. R. de Sciences \'Economiques, Gestion, Math\'ematiques et
Informatique, \\ CNRS UMR7536 and Universit\'e Paris X-Nanterre,
92001 Nanterre Cedex, France \\
$^2$ Laboratoire de Physique de la Mati\`{e}re Condens\'{e}e\\ CNRS UMR6622 and
Universit\'{e} de Nice-Sophia Antipolis, 06108 Nice Cedex 2, France\\
$^3$ Institute of Geophysics and
Planetary Physics and Department of Earth and Space Science\\
University of California, Los Angeles, California 90095\\
e-mails: vitting@unice.fr and sornette@unice.fr\\}

\date{\today}
\maketitle

{\bf
We propose a payoff function extending Minority Games (MG) that captures
the competition between agents to make money. In constrast with previous MG, the best
strategies are not always targeting the minority but are shifting opportunistically
between the minority and the majority. The emergent properties of the price
dynamics and of the wealth of agents are strikingly different from
those found in MG. As the memory of agents is increased, we find a phase
transition between a self-sustained speculative phase in which 
a ``stubborn majority'' of agents effectively 
collaborate to arbitrage a market-maker for their mutual benefit and 
a phase where the market-maker always arbitrages the agents. A subset
of agents exhibit a sustained non-equilibrium risk-return profile.
}

\vskip 0.5cm
\vskip 0.5cm

The Minority Game (MG)\cite{minoritygames-standard} is perhaps the simplest
in the class of multi-agent games of interacting
inductive agents with limited abilities competing for scarce resources. 
Many published works on MG have motivated their study by their relevance to
financial markets, because investors exhibit a large heterogeneity of 
investment strategies, investment horizons, risk aversions and wealths,
have limited resources and time to dedicate to novel strategies and
the minority mechanism is found in markets. Here, our goal is to point out
that the minority mechanism is a relatively minor contribution to the 
self-organization of financial markets. We develop a better description
based on a financially motivated payoff function.
Following the standard specification of MG, we assume that markets
are purely speculative, that is, agents profit only from changes 
in the stock price. In addition,
agents are chartists or technical analysts who only analyze past realization
of prices, with no anchor on fundamental economic analysis. 

A MG is a repeated game where $N$ players have to choose one
out of two alternatives at each time step 
based on information represented as a binary time series $B(t)$. Those
who happen to be in the minority win.
Each agent $i$ possesses a memory of the last $m$ digits of $B(t)$.
A strategy gives a prediction for the next
outcome of $B(t)$ based on the history of the last $m$
digits of $B$. Since there are $2^m$ possible histories, the total number
of strategies is given by $S=2^{2^m}$. Each agent holds the same number $s$ of
(but in general different) strategies among the $S$ possible strategies. 
At each time $t$, every agent uses her most successful strategy (in
terms of payoff, see below) to decide whether to buy or sell
an asset. The agent takes
an action $a_i(t) = \pm 1$ where $1$ is interpreted as
buying an asset and $-1$ as selling an asset. The excess demand, 
$A(t)$, at time $t$ is
therefore given as $A(t) = \sum_{i=1}^{N} a_i(t)$.
The payoff of agent $i$ in the MG is given by:
\be
g_i(t) = - a_i(t) A(t)  ~.  \label{MG-payoff}
\ee
As the name of the game indicates, if a strategy $i$ is in the minority
($a_i(t)A(t) < 0$), it is rewarded. In other words, agents in MG try to be
anti-imitative. To ensure causality, 
the notation $- a_i(t) A(t)$ in  (\ref{MG-payoff})  
must be understood as $ - a_i(t-1/2) A(t)$
since the actions/strategies of the agents take place {\em before} the
price (and thus the payoff) can be determined. 
The richness and complexity of 
minority games stem from the fact that agents have to be different; 
theories based on an effective 
representative agent are bound to fail because she would represent the majority. MG 
are intrinsically frustrated and fluctuations and heterogeneities are the key 
ingredients.

In order to model financial markets, several 
authors have used
the following or slight variants of the following equation
for the return $r(t)$ \cite{bouchaudcont,farmer}
\be
r(t) \equiv \ln (p(t+1)) - \ln (p(t)) =  A(t)/\lambda ,  \label{price-eq}
\ee
where $\lambda \propto N$ is the liquidity. The fact that the price
goes in the direction of the sign of the order imbalance $A(t)$ is 
well-documented
\cite{Holthausen,Lakonishok,Chan,Maslov,ChalletStin}.
By constructing and analyzing a large
database of estimated market-wide order imbalances for a
comprehensive sample of NYSE stocks during the period 1988-1998 inclusive,
Chordia et al. \cite{Chordia} confirm that contemporaneous order 
imbalance $A(t)$
exerts an extremely significant impact on market returns in the expected
direction; the positive coefficients of their regressions
imply that excess buy (sell) orders drive up (down) prices,
in qualitative agreement with (\ref{price-eq}).

Let us assume that an agent thinks at time $t-1/2$
that the unknown future price $p(t)$ will be larger than the known
previous quote $p(t-1)$ and larger
than the next future quote $p(t+1)$, thus identifying $p(t)$ as a local maximum.
Her best strategy is to put a
sell order at time $t-1/2$ in order for the sale to be realized at time
$t$ at the local price maximum, allowing her to profit from future drops
at later times. She will then profit and cash in the money equal to the drop
from the local maximum at time $t$ to a smaller price realized at $t+1$ 
or later. In this case, the optimal strategy
is thus to be in the minority as seen from the relation between the
direction of the price change given by the sign of $r(t)$ and the direction of the majority
given by the sign of $A(t)$. Alternatively, if the
agent thinks at time $t-1/2$ that $p(t-1) < p(t) < p(t+1)$, her best strategy is
to put a buy order at time $t-1/2$, realized at the price $p(t)$ at
time $t$. She will then profit by the amount $p(t+1)-p(t)$
if her expectation that $p(t) < p(t+1)$ is born out.
In this case, it is profitable for an agent to be in the majority, because
the price continues to go up, driven by the majority, as seen from (\ref{price-eq}).
In order to know when the price reaches its next local extremum
and optimize their gains, the agents
need to predict the price movement over the next {\bf two} 
time steps ahead ($t$ and $t+1$),
and not only over the next time step as in the standard MG. This pinpoints
the fundamental misconception of MG as models of financial markets.
Indeed, by shifting from minority to majority strategies and vice-versa,
an agent tries at each time step to gain $|p(t+1)-p(t)|$
whatever the sign of $p(t+1)-p(t)$: an ideal strategy is a ``return rectifier.''
Because an agent's decision $a(t-1/2)$ at time $t-1/2$ is put into practice
and invested in the stock market at time $t$, the decision will bring its
fruit from the price variation from $t$ to $t+1$. From
(\ref{price-eq}), this price variation is simply proportional to $A(t)$.
Therefore, the agent has a positive payoff if $a(t-1/2)$ and $A(t+1/2)$ 
have the same sign.
As a consequence, in the spirit of the MG (and using the MG notation without 
half-time scales), the correct payoff function is:
\be
g^\$_i(t+1) =  a_i(t) A(t+1) ~.  \label{our-payoff}
\ee
The superscript $\$$ is a reminder that the action taken by
agent $i$ at time $t$ results at time $t+1$ in a percentage
gain/loss of $g^\$_i(t+1)/\lambda$ (see (\ref{price-eq})). We will
refer to the game where the agents use (\ref{our-payoff}) as the
``\$-game'' since, by using this payoff function, the agents strive
to increase their wealth. This reasoning stresses that,
in real markets, the driving force underlying the competition between investors 
is not a struggle to be in the minority at each time step, but rather a fierce 
competition to gain money. 

In the simplest version of the model, 
each trade made by an agent is the exchange of one quanta of
a riskless asset (cash) for one quanta of a risky one (asset)
irrespective of the agent's wealth or the price of the asset. The
wealth of the i'th agent at time $t$ is given as
\be
W_i(t) = N_i(t) p(t) + C_i(t) ~,   \label{wealth}
\ee
where $N_i(t)$ is the number of assets held by agent $i$ and
$C_i(t)$ the cash possessed by agent $i$ at time $t$. 
In order to illustrate the differences between the payoff functions
(\ref{MG-payoff}) and (\ref{our-payoff}), we have plotted in Fig.~1 an example
of the
payoff (upper plot) of the best as well as the worst performing MG agent using
(\ref{MG-payoff}). Each agent is allowed to take either a long or
a short position, and we furthermore assume that the agents stay
in the market at all times. This means that if e.g. an agent has
taken a long position (i.e. taken the action $a_i=1$ to buy a asset)
the agent will not open new positions (and therefore does not contribute
to the excess demand and price change) but keep the long position
until she gets a signal to sell ($a_i=-1$)\cite{note2}.
The lower plot of Fig.~1 shows the wealth (\ref{wealth})
corresponding to the agents
of the upper plot.
The consistently bad performance of the optimal MG-agent in terms of her wealth and
reciprocally the relatively good performance for the worst MG-agent in terms
of her wealth is a clear illustration of the fact that a minority strategy 
will perform poorly in a real market. This does not exclude howeer the potential
usefulness of MG strategies in certain situations, in particular
for identifying extrema, as discussed above and
as illustrated recently in the prediction of 
extreme events \cite{Johnson-Lamper-Jefferies-Hart-Howision}.
In contrast, for the ``$\$$-game'' (\ref{our-payoff}) presented here,
the performance of the  payoff function (\ref{our-payoff})
matches by definition the performance of the wealth of the 
agents. The superficial observance by some MG of the 
stylized facts of financial time series is not a proof of their relevance
and, in our opinion, express only the often observed fact that many models,
be they relevant or irrelevant, can reproduce superficially a set of 
characteristics (see for instance a related discussion on mechanisms 
of power laws and self-organized criticality in chapters 14 and 15 of
\cite{DSbook}).

In order for trading to occur and to fully specify the price trajectory, 
a clearing mechanism has to be specificed. Here, we use a 
market maker who furnishes assets in
case of demand and buys assets in case of
supply \cite{Jefferies-Hart-Hui-Johnson}. 
The price fixing equation (\ref{price-eq}) implicitely assumes
the presence of a market-maker, since the excess demand of the agents $A(t)$ always 
finds a counterpart. For instance, if the cumulative action of the agents is to 
sell 10 stocks, $A(t)=-10$, the
market-maker is automatically willing to buy 
10 stocks at the price given by (\ref{price-eq}).
As pointed out in Ref.\cite{Jefferies-Hart-Hui-Johnson}, 
expression (\ref{price-eq}) leads to an unbound market-maker inventory
$S_M(t)$. In order to lower his inventory costs and the risk
of being arbitraged, a market-maker will try 
to keep his inventory secret and in average close to zero \cite{Chordia}. As 
shown in \cite{Jefferies-Hart-Hui-Johnson}, this can be achieved by the 
following generalization of (\ref{price-eq}): 
\be
r(t) \equiv \ln (p(t+1)) - \ln (p(t)) =  (A(t)-S_M(t))/\lambda ,  \label{price-eq-with-S}
\ee
with $S_M(t) = -\sum_{t=0}^{t-1} A(t)$. Expression (\ref{price-eq-with-S}) implies that, 
the larger is the long position the market-maker is holding, the more he will lower
the price in order to attract buyers, and vice-versa for a short 
position. Another way to ensure the same behavior is to 
introduce a spread or change the available liquidity \cite{note3}.

We first study the price formation using (\ref{price-eq}) and resulting from a 
market competition between agents with payoff 
function (\ref{our-payoff}) and compare it with the MG case (\ref{MG-payoff}) 
in the case with no constraint on the number of stocks held by
each agent (i.e., an agent can open a new position at each 
time step). Contrary to the MG case, we find that the price always  
diverges to infinity or goes zero within a few tens or hundreds of time steps. 
This behavior is observed for all values of $N, m, s$. Similar results are
found if we replaced the price equation (\ref{price-eq}) with (\ref{price-eq-with-S})
which includes the market-maker strategy. The reason for 
this non-stationary behavior is that agents, using (\ref{our-payoff}) as 
pay-off function, are able to collaborate to their mutual benefit.  This 
happens whenever a majority among the agents can agree to ``lock on'' 
for an extended period of time to a common decision of either to keep on 
selling or buying. A constant sign of $A(t)$ is seen from either (\ref{price-eq}) 
-(\ref{wealth}) or (\ref{our-payoff})-(\ref{price-eq-with-S})
to lead to a 
steady increase of the wealth of those agents sticking to the majority 
decision. A ``stubborn majority'' manages to collaborate by 
sticking to the same common decision - they all gain by doing so at the 
cost of the market-maker who is arbitraged. The mechanism underlying
this cooperative behavior is the positive feedback
resulting from a positive majority
$A(t)$ which leads to an increase in the price 
(\ref{price-eq-with-S}) which in turn confirms the ``stubborn majority'' 
to stick to their decision and keep on buying, leading to a further
confirmation of a positive $A(t)$. This situation is reminiscent of 
wild speculative phases in markets, such as occurred prior to the October 1929
crash in the US,  before the 1994 emergent market crises in Asia, and more
recently during the ``new economy'' boom on the Nasdaq stock exchange,
in which margin requirements are decreased and/or investors are allowed to borrow
more and more on their unrealized market gains. 
This situation is quite parallel to
our model behavior in which agents can buy without restrain, pushing the prices
up. Of course, some limiting process will eventually appear, often leading to 
the catastrophic stop of such euphoric phase.

We turn to the more realistic case where agents have bounded 
wealth, and study the limiting case where agents are allowed to keep 
only one long/short 
position at each time step. With this constraint, the previous
positive feedback is no longer at work. Holding a position, an agent
will contribute to future price changes only
when she changes her mind. Thus, a ``stubborn
majority'' can not longer directly influence future price changes through the 
majority term $A(t)$, but only now indirectly through the impact on the
market maker strategy $S_M(t)$ in (\ref{price-eq-with-S}). 
Fig.~2a show typical examples of price trajectories using
(\ref{our-payoff})-(\ref{price-eq-with-S}) with agents keeping a single position
(short/long) at any times, for three different choices of parameter 
values $(N,m,s)$. The time series are quite similar to typical financial price series
and possess their basic stylized properties (short-range
correlation of returns, distribution of returns with fat tails, long-range
correlation of volatility).
The corresponding wealth of the market maker is shown in Fig.~2b. It exhibits
a systematic growth, interrupted rarely for some short periods of 
time with small losses. The stochastic nature of the price
trajectories is translated into an almost deterministic wealth growth
for the market-maker, who is an almost certain winner (as it should and is
in real market situations to ensure his survival and profitability). The market
maker is similar to a casino providing services or entertainments and
which profits from a systematic bias here resulting from
the lack of cooperativity of the agents.

For each agent $i$, we define a risk parameter
\be
R_i(t) = \langle (dW_i(t)- \langle dW_i \rangle _t)^2 \rangle_t
\label{risk-return-eq}
\ee
where $dW_i(t)$ is the change of wealth of agent $i$ between $t$ and $t-1$.
$R_i(t)$ is the volatility of the wealth of agent $i$.
The average return per time step $\langle dW_i \rangle$ for 
each of the $N=101$ different agents as a function of his volatility $R_i$ 
is shown in Fig.~2c (each point corresponds to one agent).
Since agents choose either a short or a long position at each time step, a
perfect performing agent is a return rectifier taking no risk. Similarly, the
worst performing agent is consistently moving against the market, again with
the risk defined from (\ref{risk-return-eq}) equal to zero. 
This explains why the 
risk-return behavior seen in Fig.2~c is an mirror image of the risk-return
efficient frontier in Markovitz standard portfolio theory \cite{Markovitz}.
The figure shows that even though the market-maker arbitrages the agents
as a group, some ``clever'' agents are still able to profit from  their
trade with a risk-return profile which should be unstable
in the sense of standard economic theory. It is however a robust and
stable feature of our model. 
This property results fundamentally from the heterogeneity of the
strategies and can not be captured by a representative agent theory.

To study further the competition between the agents as a group and the 
market-maker, we let the \$-game evolve for $T$ time steps and measure if
the market-maker has arbitraged the agents, i.e., if his wealth is positive
at the end of the time period $T$. Fig.~3 shows the
probability $P(m)$ for the market-maker to arbitrage the agents
versus the memory of the agents $m$. For $m=1$, the agents 
always exploit the market-maker according to the positive feedback
mechanism involving the ``stubborn majority'' described above.
As $m$ increases, $P(m)$ increases and, 
for the largest memory $m=11$ of the agents, the
market-maker arbitrages the group of agents with probability $1$. 
This correspond to the
examples illustrated in Fig.~2. In between, there is a competition between
cooperativity between the agents and the destructive interferences of
their heterogeneous strategies.
The finite-size study of $P(m)$ as a function of $T$ suggests the 
existence of a sharp transition in the large $T$ limit for $m \approx 9$.
Below this memory length, the set of strategies available to agents
allow them to sometimes cooperate successfully. As the complexity of the
information increases, their strategies are unable to cope with the
large set of incoming information and the chaotic desynchronized behavior
that results favors the market maker. This could be termed the curse of
intelligence. 

We will report elsewhere on extensions of this model  
with traders who act at different time scales and with different 
weights and on the detection of large price movements in the spirit of 
\cite{Johnson-Lamper-Jefferies-Hart-Howision}. 

D.S. gratefully acknowledges
support from the James S. McDonnell Foundation 21st Century Scientist 
award/studying complex systems.

\begin{figure}[h]
\includegraphics[width=8cm]{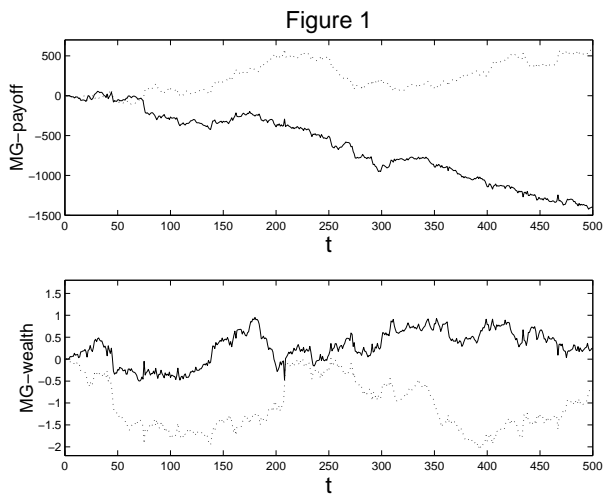}
\caption{\protect\label{Fig1} Payoff function (\ref{MG-payoff}) (upper graph)
and wealth (lower graph) for the MG-game showing the best (dotted line) and
worst (solid line) performing agent for a game using  $N=501$ agents,  memory of $m=10$ and
$s=10$ strategies per agent.  No transaction costs are applied.
}
\end{figure}

\begin{figure}[h]
\includegraphics[width=8cm]{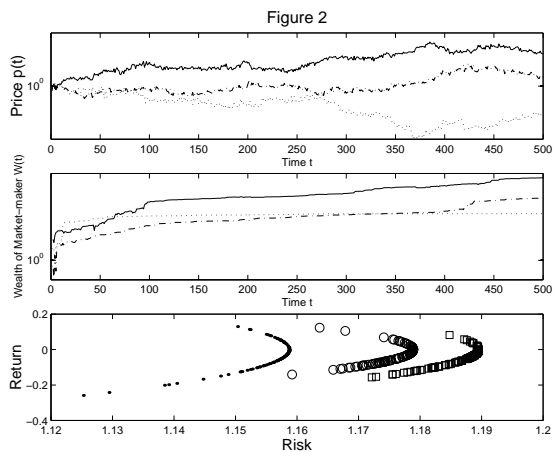}
\caption{\protect\label{Fig2} Price, wealth of market-maker and risk-return
plots for three different parameter choices using the payoff function (\ref{our-payoff})
and the constraint that agents can only accumulate one position at a time. 
Solid line and black circle: $m=10, s=4$; dashed-dotted line and circle: $m=10,
s=10$; dotted line and square: $m=8, s=10$. 
}
\end{figure}

\begin{figure}[h]
\includegraphics[width=8cm]{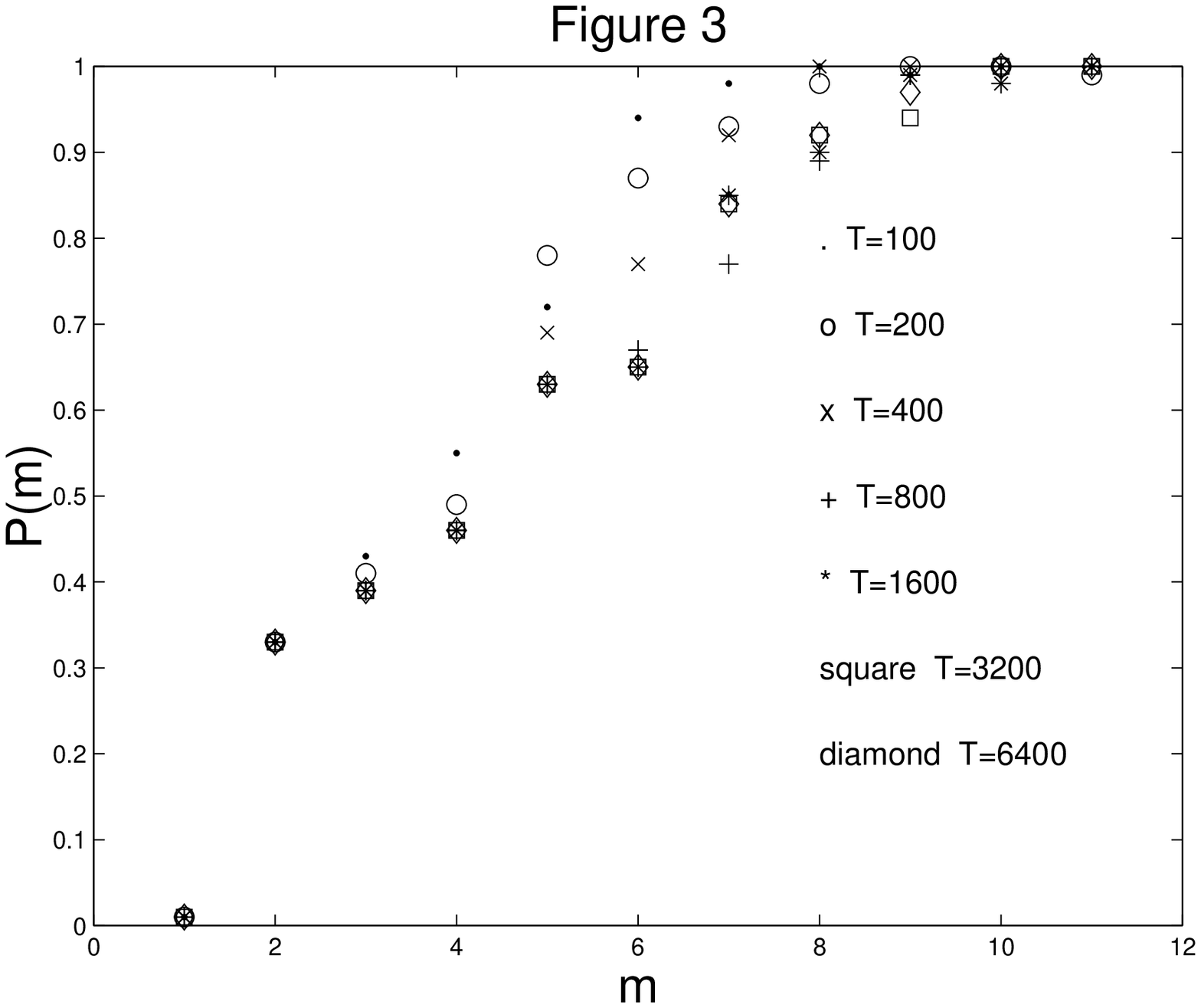}
\caption{\protect\label{Fig3} Probability $P(m)$ for the market-maker to arbitrage 
the group of agents using (\ref{our-payoff})-(\ref{price-eq-with-S}) as 
a function of the memory length
$m$. $P(m)$ is determined from the market-maker wealth after $T$ time steps and by
averaging over 100 simulations with different
initial configurations. The parameters used are
$N=101,  s=5$. Similar results are found using different $N,s$.
}
\end{figure}


\begin{thebibliography}{}
\bibitem{note2} A similar conclusion as seen in Fig.~1 is found 
in the absence of constraint on the number of stocks each agent is  
allowed to hold, that is, when agents are allowed to open a new 
position at each time step so as to leverage their strategy.
\bibitem{note3} We do not expect any qualitative 
difference in the results represented in this paper depending on which of 
these different mechanisms is implemented, since 
a ``stubborn majority'' can manifest itself independent 
of which of these mechanisms is used.
\end{thebibliography}
\end{document}